\documentclass[twocolumn]{aastex61}
\pdfoutput=1 
\usepackage{amsmath,amstext}
\usepackage[T1]{fontenc}
\usepackage{apjfonts} 
\usepackage{bbding}
\usepackage[figure,figure*]{hypcap}

\newcommand{\mgii}{Mg {\small II} h\&k}
\newcommand{\mgiik}{Mg {\small II} k}
\newcommand{\mguv}{Mg  {\small II} UV triplet}
\newcommand{\mguvtt}{Mg  {\small II} UV2\&3}
\newcommand{\mguvo}{Mg  {\small II} UV1}
\newcommand{\cii}{C {\small II} 1334 \& 1335 \AA}

\newcommand{\vturb}{$v_{turb}$}
\newcommand{\vlos}{$v_{los}$}
\newcommand{\nne}{$n_e$}
\newcommand{\ltau}{$log(\tau)$}

\newcommand{\irissq}{$IRIS^{2}$}
\newcommand{\irissqp}{$IRIS^{2+}$}
\newcommand{\ifns}{{\footnotesize I}}
\newcommand{\iifns}{{\footnotesize II}}
\shorttitle{$IRIS^{2+}$ database for active region models}
\shortauthors{Sainz Dalda et al.}

\begin{document}


\title{$IRIS^{2+}$: A comprehensive database of stratified thermodynamic models in the low solar atmosphere.}  

\author[0000-0002-3234-3070]{Alberto Sainz Dalda}
\affil{Lockheed Martin Solar \& Astrophysics Laboratory, 3251 Hanover Street, Palo Alto, CA 94304, USA}
\affil{Bay Area Environmental Research Institute, NASA Research Park, Moffett Field, CA 94035, USA.}
\author{Aaryan Agrawal}
\affil{Henry M. Gunn High School, 780 Arastradero Rd, Palo Alto, CA 94306}
\author[0000-0002-8370-952X]{Bart De Pontieu}
\affil{Lockheed Martin Solar \& Astrophysics Laboratory, 3251 Hanover Street, Palo Alto, CA 94304, USA}
\affil{Rosseland Center for Solar Physics, University of Oslo, P.O. Box 1029 Blindern, NO-0315 Oslo, Norway}
\affil{Institute of Theoretical Astrophysics, University of Oslo, P.O. Box 1029 Blindern, NO-0315 Oslo, Norway}
\author[0000-0002-5879-4371]{M.~Go\v{s}i\'{c}}
\affil{Lockheed Martin Solar \& Astrophysics Laboratory, 3251 Hanover Street, Palo Alto, CA 94304, USA}
\affil{Bay Area Environmental Research Institute, NASA Research Park, Moffett Field, CA 94035, USA.}

\begin{abstract}
Our knowledge of the low solar atmosphere, i.e., the photosphere and chromosphere, is based on the knowledge gained from the observations and the theoretical and  numerical modeling of these layers. In this sense, the thermodynamical and magnetic semi-empirical models of the solar atmosphere have significantly contributed to the advance in the understanding of the physics of the Sun. In the past, many of these models have been used as a reference that helps us to, e.g., constrain the theoretical and numerical modeling, or to verify the goodness of physical parameters obtained from the inversion of the spectral lines. Nevertheless, semi-empirical models are quite limited by the assumptions that are inherent to the approach and do not necessarily provide an accurate view of the instantaneous and local thermodynamic conditions in the solar atmosphere. In this work, we provide an extensive collection of thermodynamic model atmospheres for active regions (ARs) obtained from the
simultaneous inversion of 6 lines sensitive to changes in the thermodynamical conditions in
the chromosphere, and another 6 lines sensitive to changes in the thermodynamical conditions
in the photospere. These inversions were made using 320  {\it representative
profiles} (RP) obtained by clustering the profiles in the umbra, penumbra, pore-like, plage, and surrounding quiet-sun in 126 active regions. Due to the simultaneous inversion of the selected lines, the resulting {\it representative model atmosphere} (RMA) samples the thermodynamics from the bottom of the photosphere to the top of the chromosphere. 
As a result, this database, named \irissqp\ and formed by $40,320$ RP-RMA pairs, represents 
the most comprehensive collection of stratified-in-optical-depth thermodynamic models of the low solar atmosphere.  
\end{abstract}


\section{Introduction}

Modern astrophysics started when we were able to derive physical information - beyond morphological and
phenomenological behavior - from the objects we observed in the sky.  We can establish the last half of the $19^{th}$ century as the birth of what we now call astrophysics. Two critical techniques were needed: 
spectroscopy and photography, i.e.: the capability to record the light coming from the Sun (or any other object in the sky) as a function of wavelength. The comparison of data thus obtained with similar data obtained in the laboratory has led to great progress in understanding the physical conditions on stars.
Key to the interpretation of these new measurements was the development of models that led to a greater understanding of 
matter, i.e., the atomic model, and light, more
precisely, electromagnetic radiation. A quintessential example of this
synergy between experiments, techniques and theory is the discovery of the
magnetic field in the Sun(spots), inferred through the Zeeman effect in the
spectral lines observed on sunspots by \citet{Hale08a}. Since then, many
advances, both theoretical and technological have improved our knowledge about
the matter, the electromagnetic radiation, and the physics of the objects
observed in the sky.

The study of the Sun has played a prominent role in the knowledge we
now have about stars, and in the development of the techniques and ideas
needed to gain that knowledge. Thanks to its proximity, we are
able to resolve many structures on its surface (photosphere), and
the rest of the solar atmosphere, i.e., the chromosphere \citep{Rutten07,Cauzzi08b,Wedemeyer16,Carlsson19}, transition region, and the corona. 
Current instrumentation allows us to recover in great detail the physical
conditions of solar structures as small as $100~km$, with a temporal cadence of
a few seconds, and with a high sensitivity both in the magnetic field
($\delta|B|<10~G$) and in other derived thermodynamic parameters (e.g., $\delta T <50~K$).

There are two main ways to gain knowledge about the solar atmosphere. Forward modeling involves numerical models of the thermodynamic (or magnetothermodynamic) structure and comparing the synthetic spectral lines or {\it profiles} obtained by solving the radiative transfer equation (RTE) with solar observations. This method is not unique as it starts with an initial guess at the thermodynamic parameters and assumptions about the dominant physical processes.
The second method recovers a model atmosphere from iteratively fitting synthetic profiles - obtained by solving the RTE from a initial guess model atmosphere and slight subsequent modifications -  until a best fit is found between the observed profile and the synthetic profile. This approach is also not unique as it sensitively depends on the initial guess and could suffer from common issues with least-square fitting such as local minima in the loss function. Because of that, we (should) say that {\it the recovered model
atmosphere is compatible with the observed profile}, and therefore that these 
physical conditions {\it may occur} in the solar atmosphere. This last tool is
called the inversion of spectro(polarimetric) data\footnote{The reader
interested in a detailed review of this method can be found in
\citet{delToroIniesta16}}. 

While neither method provides a unique solution, both can provide valuable insight into the physical conditions and processes that dominate the solar atmosphere. For both methods it is important to have an accurate knowledge of the model of the solar atmosphere. In the forward modeling, this knowledge comes from the physics recovered from previous observations and by the theoretical considerations of the problem (e.g., hydrostatic equilibrium, ambipolar difusion, \citealt{Khomenko14,Martinez-Sykora20}), which help us constrain the physical conditions considered in the model. For the inversion, the feasibility of the initial guess model used to start the inversion is critical, as well as the theoretical considerations made while solving the RTE (e.g., statistical equilibrium, non-local thermodynamic equilibrium, partial frequency distribution, e.g.,  \citealt{Leenaarts13a}), and the atom model used to synthesize the spectral lines. In fact, the theoretical considerations should be similar in both cases, since both methods try to reproduce the observables we detect, i.e., the spectral line profiles. Both methods suffer from computational and algorithmic limitations, whether it is the inclusion of the effects of multi-fluid interactions (forward modeling, see \citealt{Wargnier22}) or the calculation of 3D radiative transfer in chromospheric lines (inversions, see \citealt{Sukhorukov17,Bjorgen19}). 


In this paper, we focus on providing a collection of
thermodynamic model atmospheres for active regions (ARs). We consider an active
region as the non-exclusive combination of umbra, pores or pore-like structures, penumbra, plage and the 
neighboring quiet Sun. There are many solar atmosphere models available\footnote{An excellent set of solar atmosphere models has been collected by Dr. Basilio Ruiz Cobo. They are available at \href{https://github.com/BasilioRuiz/SIR-code/tree/master/models}{https://github.com/BasilioRuiz/SIR-code/tree/master/models}.}, most of them are particularly devoted to one solar feature, e.g., the HSRA quiet-sun model \citep{Gingerich71}, a hot sunspot model \citep{Collados94}, a penumbra model \citep{delToroIniesta94}, or a plage model \citep[Model1004][]{Fontenla09}. Some of these models are considered as the standard model for their corresponding features. In the case of the inversions, these models are mostly used as initial guess models.
The results presented in this paper complement these models by providing a set of $40,320$ {\it representative model atmosphere} (RMA) and their corresponding RPs, which are labeled with the feature where they were observed, e.g. ``{\tt plage}'', their location on the solar disk, and the recording date. Therefore, a more diverse set of models and profile are given, and they can be used to constrain numerical models, as initial model guesses for inversions, or for the synthesis of profiles at different wavelengths. 

In Section \ref{sec:data} we detail the main characteristics of the IRIS data that we have used in our study, which spectral lines we have selected for the inversions, and where in the solar atmosphere these lines are sensitive to changes in the thermodynamic conditions. We explain how we have selected the different areas in the AR data, i.e. umbra, pore-like, penumbra, plage, and surrounding quiet Sun. We justify the need to cluster these data and how we treated the spectral lines to be simultaneously inverted. In Section \ref{sec:results} we show the results of the inversions, i.e. the fit between the observed profiles and the synthetic profile provide by the inversion, as well as the model recovered from that inversion. As a result, we present a database of synthetic RPs and RMAs in the low solar atmosphere. This data base is called \irissqp, since it represents an extension - in terms of the optical depth where the models are now more accurate - to \irissq. In Section \ref{sec:discussion} we discuss the confidence and the usability of \irissqp\ database.

\section{Data}\label{sec:data}
\begin{figure*}
    \centering
    \includegraphics[width=.3\textwidth]{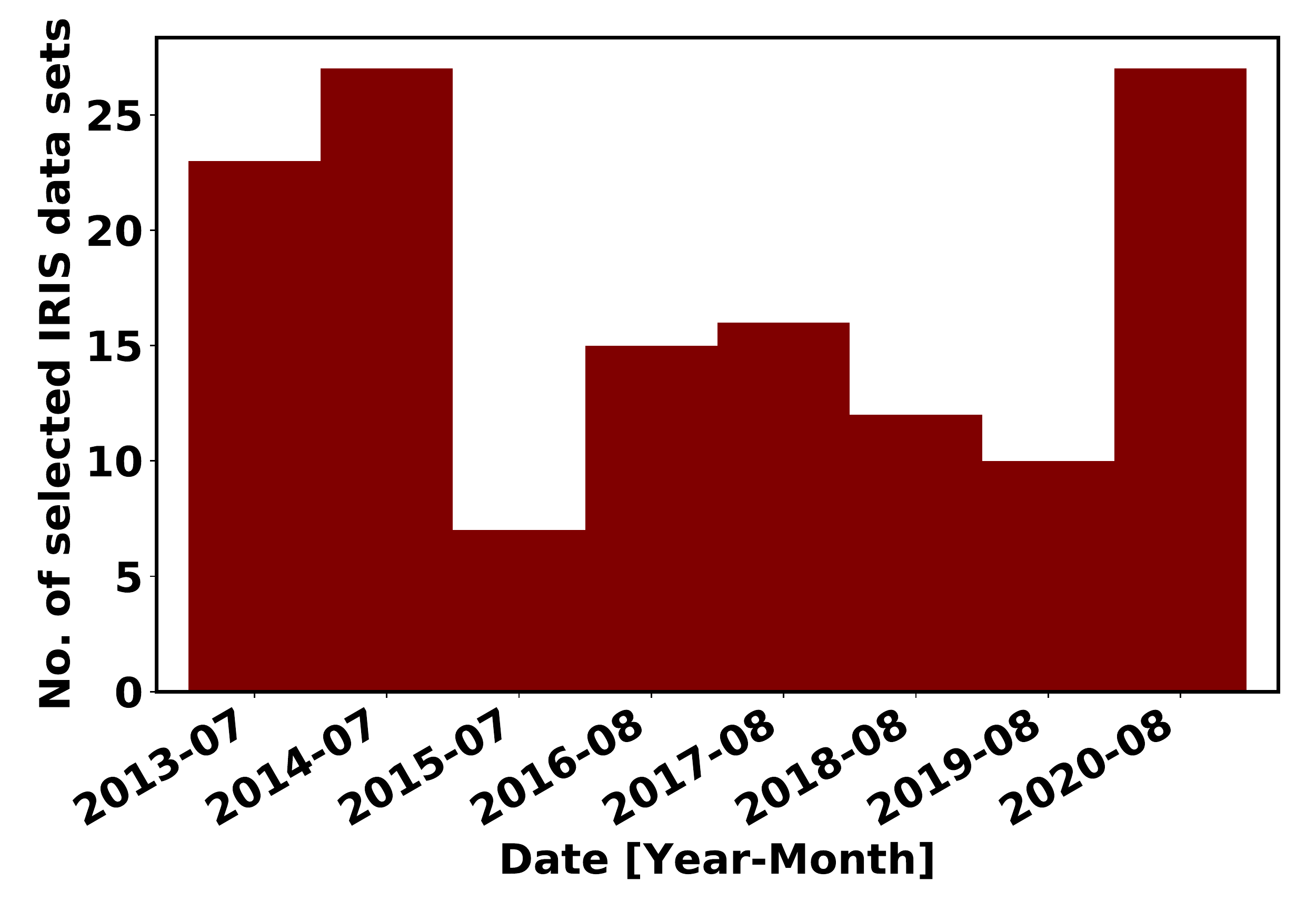}
    \includegraphics[width=.3\textwidth]{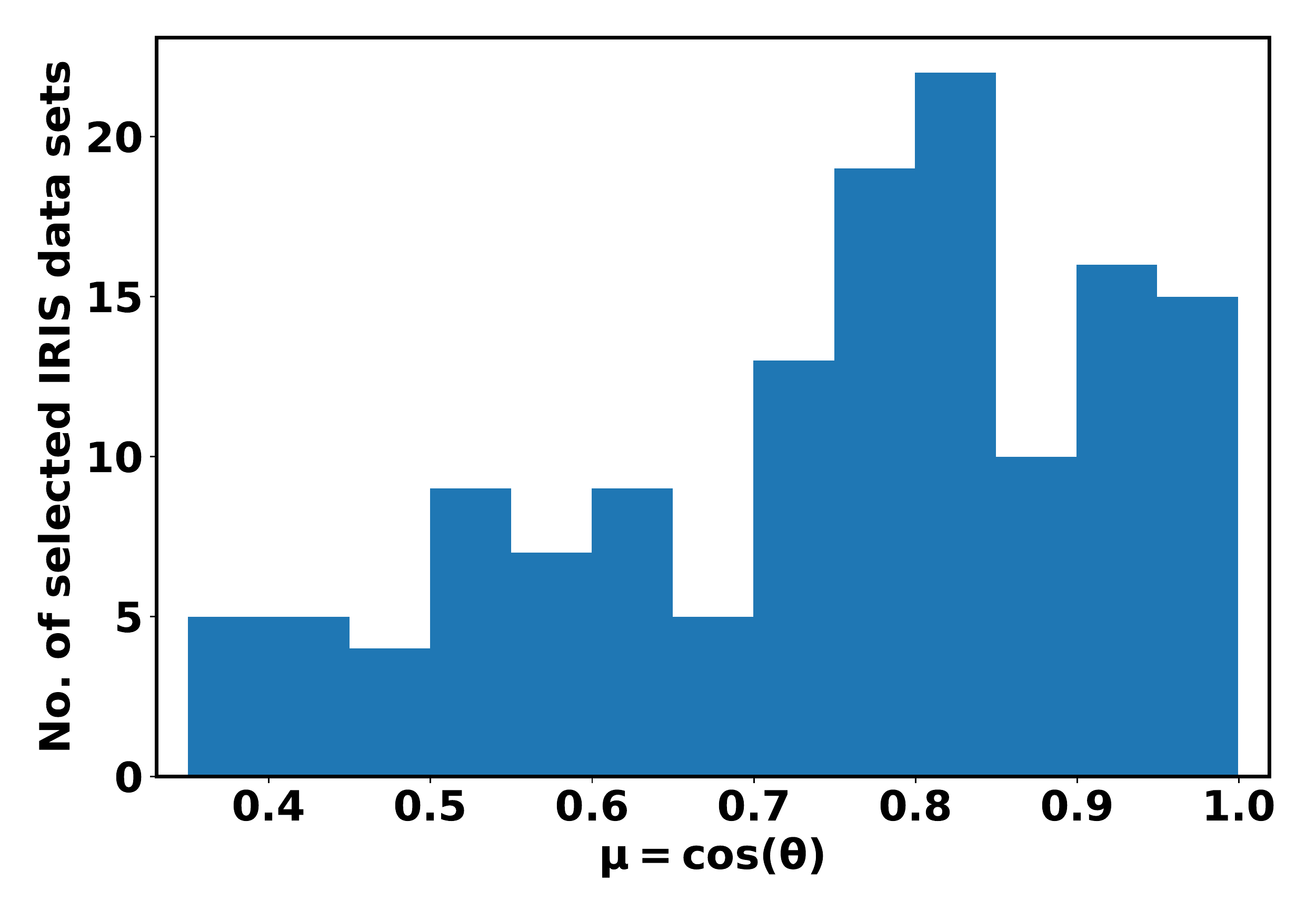}
    \includegraphics[width=.3\textwidth]{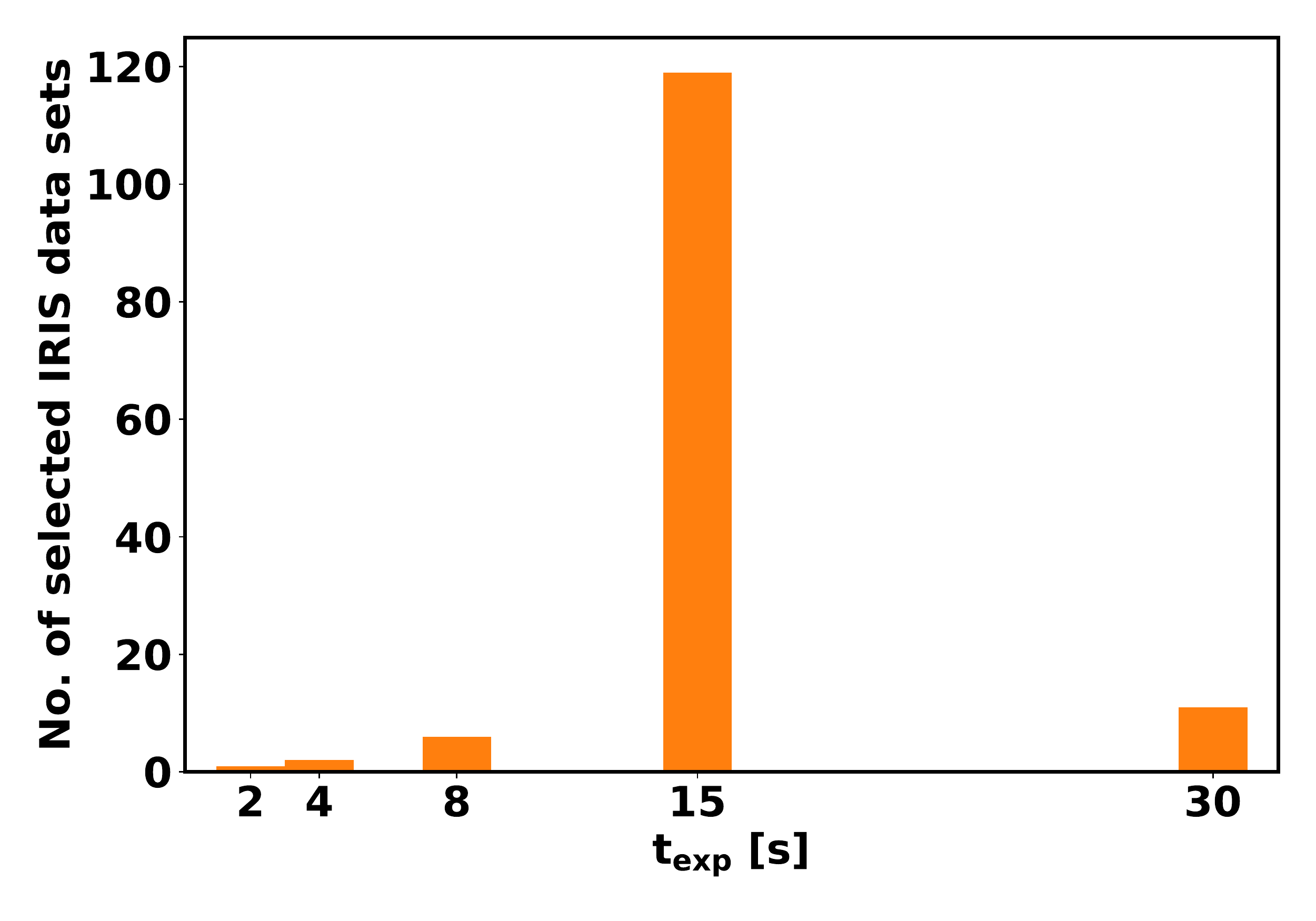}
    \caption{Distribution of the selected data in the \irissqp\ database with respect the observation date (left), the position in the solar disk given by $\mu=cos(\theta)$ - with $\theta$ the heliocentric angle (middle), and the exposure time (right).}\label{fig:hist}
\end{figure*}

The most critical part of this project is likely to select a set of data that represents properly the variety of active regions in the Sun. As we have already mentioned, by active region we understand the non-exclusive combination of plage, umbra, pore, and/or penumbra with surrounding quiet Sun (which may or may not be affected by the neighboring active region). Thus, an IRIS data set containing spectra from a region that combines only quiet sun and a plage region is considered in this selection as an {\it active region}. This relaxed definition does not diminish the goal of this study: to have a comprehensive collection of RPs and RMAs that characterizes the thermodynamics of the main elements of the ARs. To this aim, we have selected 126 IRIS data sets located at different positions on the solar disk, observed with different exposure times, and containing different active regions observed in the time range from July 2013 (the beginning of the IRIS mission) to 2021. Figure \ref{fig:hist} shows the distribution of the selected ARs with respect the observing date (left panel), the location on the solar disk given by $\mu=cos(\theta)$, with $\theta$ the heliocentric angle (center panel), and the exposure times (right panel). The size of the field of view (FoV) was $128\times130~arcsec^2$ for 104 data sets, $128\times175~arcsec^2$ for 14 data sets, $140\times175~arcsec^2$ for 3 data sets, $112\times124~arcsec^2$ for 3 data sets, $64\times175~arcsec^2$ for 1 data set, and $64\times124~arcsec^2$ for 1 data set. 

\subsection{Selection of spectral lines}\label{sec:sel_lines}

\begin{table*}[t]
	\begin{center}
		\begin{tabular}{c|cccccccccc}
	IRIS & Fe \ifns & Ni \iifns & C \ifns & Fe \ifns &  Mg \iifns\ h & Mg \iifns\ UV3\&2 & Mg \iifns\ k &  Fe \ifns & Mg \iifns\ UV1 & Ti \iifns \\
linelist & 2827.33 & 2815.18 & 2810.58 & 2809.15 & 2803.53 & 2798.85\&.72 & 2796.35 & 2793.22 & 2791.60 & 2785.46 \\
		\hline
v36\_00 \& v38\_00 &  &\checkmark &  &\checkmark &\checkmark &\checkmark &\checkmark &\checkmark & \checkmark &  \\
v36\_01 \& v38\_01 &  &\checkmark &  &  &\checkmark &\checkmark &\checkmark &  &  &  \\
v36\_02 \& v38\_02 &  &\checkmark &  &  &  &\checkmark &\checkmark &  &  &  \\
v36\_03 \& v38\_03 &\checkmark &\checkmark &  &  &\checkmark &\checkmark &\checkmark &\checkmark &\checkmark &  \\
v36\_04 \& v38\_04 & \checkmark &\checkmark &\checkmark &\checkmark &\checkmark &\checkmark &\checkmark &\checkmark &\checkmark &\checkmark \\
v40\_00 &  &  &  &  &\checkmark &\checkmark &\checkmark &  &  & \checkmark \\
v40\_02 &  &  &  &  &\checkmark &   &\checkmark &  &  &  \\
v40\_09 & \checkmark &\checkmark &\checkmark &\checkmark &\checkmark &\checkmark &\checkmark &\checkmark &\checkmark &\checkmark \\
		\end{tabular}
		\caption{Availability of the spectral lines (in \AA) considered in the \irissqp\ for each observation IRIS {\it linelist}. The \cii\ lines, not included in the table, are always included in any IRIS linelist, therefore, they are always considered by \irissqp.}\label{table:linelists}
	\end{center}
\end{table*}

\begin{figure*}
    \centering
    \includegraphics[width=\textwidth]{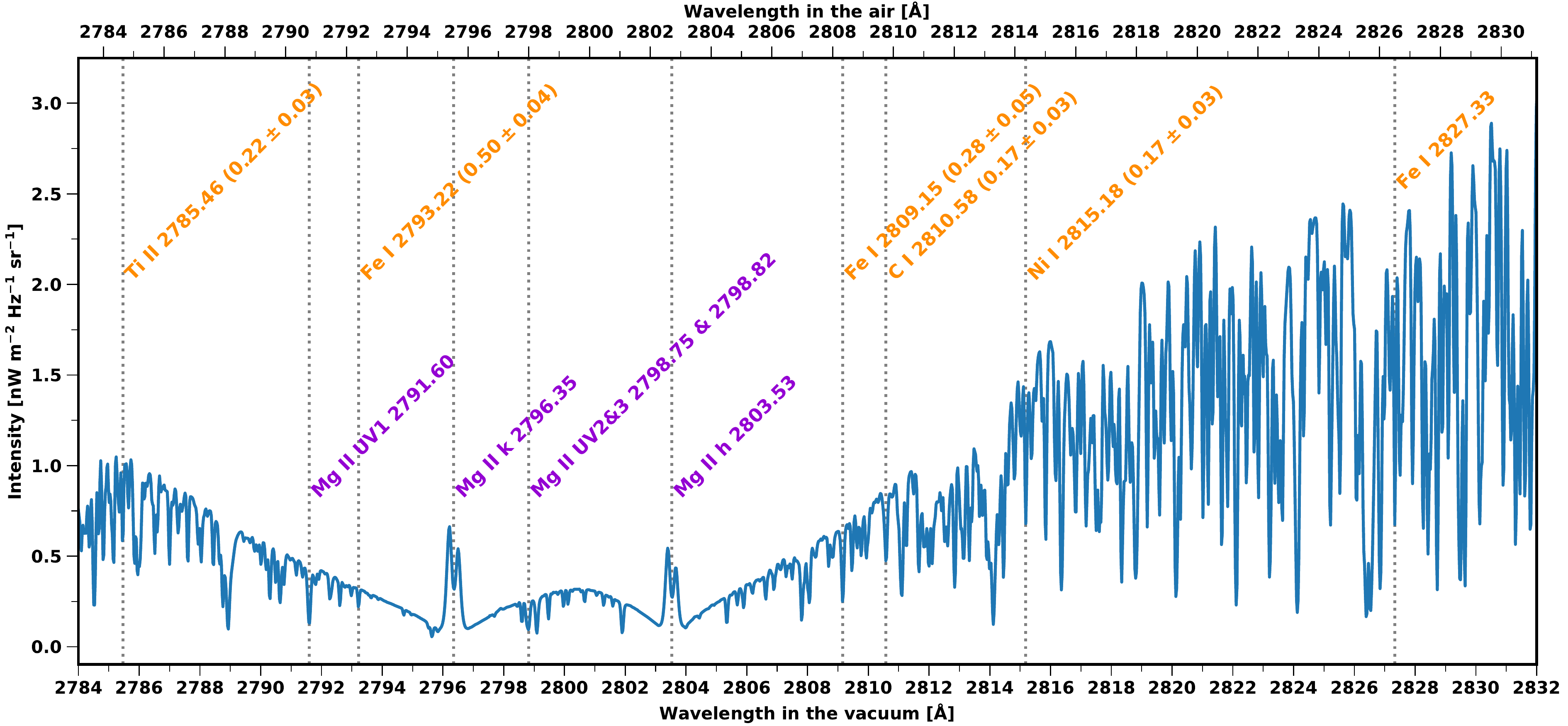}
    \includegraphics[width=\textwidth]{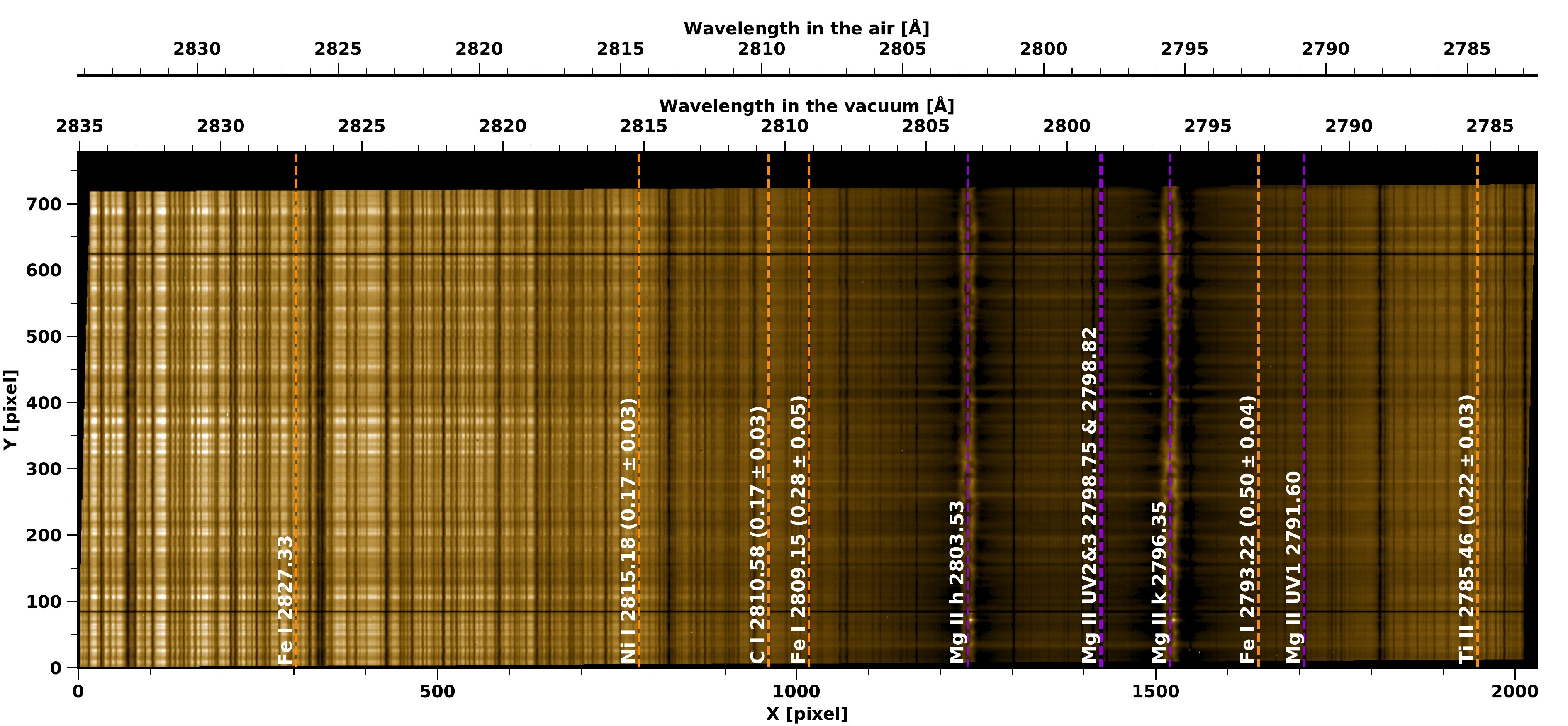}
    \caption{Top: The photospheric (orange) and chromospheric (violet) selected lines in the spectral range observed by IRIS around the \mgii\ lines. The values in parenthesis indicate the approximate height (in $Mm$) for an optical depth of 1, and the uncertainty associated with the fact that the formation occurs in a corrugated layer, as obtained by \citet{Pereira13}. Bottom: The spectrum as seen on the detector is shown for a better comparison with the IRIS {\it linelist} files, e.g. \href{https://www.lmsal.com/hek/images/iris_linelists/v36_01.png}{v36\_01}, which show all the spectral ranges selected for a given observation. Note that wavelength increases towards the left in this panel.} 
    \label{fig:speclines}
\end{figure*}

Another important issue is the selection of the photospheric spectral lines. We have tried different combinations of photospheric lines located in the near ultra-violet (NUV) spectral range observed by IRIS. The bottom panel in Figure \ref{fig:speclines} shows the full spectrum taken by IRIS in its NUV channel. In the top panel of this figure, an average profile of the quiet sun observed at the center of the solar disk is shown. The selected photospheric lines are indicated by vertical lines (orange in the bottom, black in the top 
panel), while the chromospheric lines, i.e. the \mgii\ lines and the \mguv\ lines, are similarly indicated in violet. The full spectrum along the slit that is shown in the bottom panel of this figure is intended to help the comparison with the IRIS {\it linelist} images\footnote{The IRIS {\it linelist} of an IRIS observation is shown in the "Raster" column that shows up after an observation has been selected in the right panel at  \href{https://iris.lmsal.com/search/}{https://iris.lmsal.com/search/}.} that illustrate the combination of spectral ranges selected in an IRIS observation. Because of these linelists (which are used for telemetry reasons, i.e., to minimize the data volume that is downlinked), we have to take into account the availability of the selected lines in the different IRIS observation setups or {\it linelists}. Table \ref{table:linelists} shows which selected lines are available in the IRIS {\it linelists}. Note that in our study, in addition to the lines observed in the NUV channel, we also consider the \cii\ lines recorded in the far ultraviolet (FUV) channel. 

\begin{figure*}
    \centering
    \includegraphics[width=.495\textwidth]{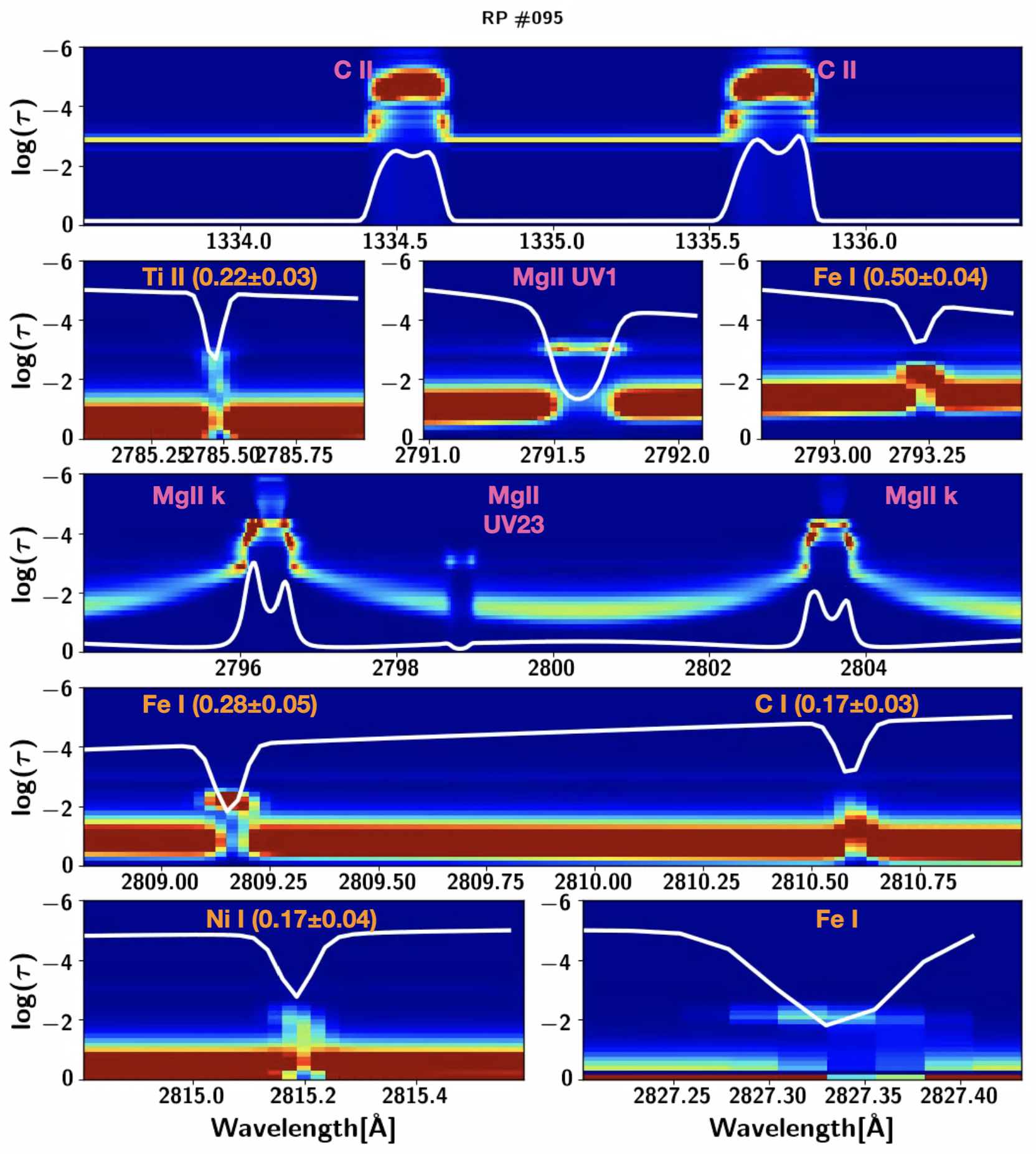}
    \raisebox{0.38\height}{\includegraphics[width=.50\textwidth]{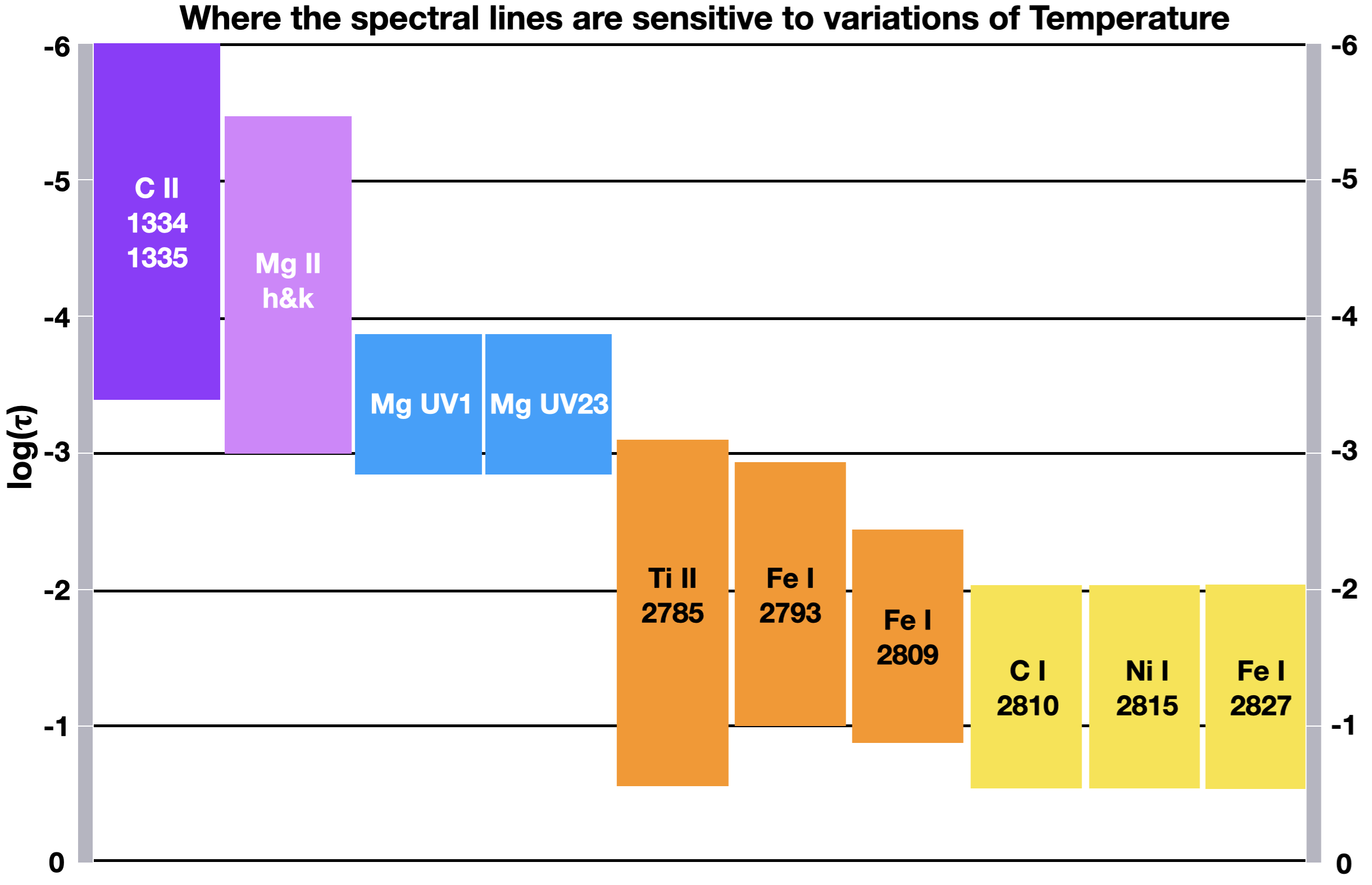}}
    \caption{Left: Response function (RF) of the intensity to variations of $T$ of the selected lines in the \irissq\ database. The spectral lines (white) are shown as a reference. Right: the optical depth ranges where these lines are on average sensitive to changes of $T$ for the solar features included in the \irissqp\ database. Note that these ranges may slightly vary for each solar feature and for the other physical parameters in the database. The values under the lines are given in \AA. The values between parenthesis are given in $Mm$. See Figure \ref{fig:speclines} for details.}\label{fig:rfs}
\end{figure*}

The selection of the photospheric lines is intended to recover as accurately as possible the variation of thermodynamics from the bottom to the top of the photosphere. This information is derived from the {\it response function} \citep[RF,][]{Mein71,LandiDegl'Innocenti81} of the intensity profiles to a variation in a given physical parameter, and from the height at which the optical depth is above 1, as obtained by \citet{Pereira13}. For all the photospheric lines in the panels of Figure \ref{fig:speclines}  - except \ion{Fe}{1} 2827.33 \AA - the  height is provided in $Mm$. including the range of heights that maps velocities in the model to observed line shifts, associated with the fact that the line formation occurs in a corrugated layer. The height estimates are based on the synthesis of these lines using a modified version of the RH code \citep{Uitenbroek01,Pereira15a} on a 3D radiative MHD numerical simulation obtained by the Bifrost code \citep{Gudiksen11} . The RFs of the intensity (for a multi-line RP in the database) to the variation of the temperature is shown in the left panel in Figure \ref{fig:rfs}. Note that the optical depth range where a spectral line is sensitive to the variation of a physical parameter is slightly different for different solar features (e.g., umbra, penumbra, filament, plage), and it is also different for the various physical parameters (temperature, electron density, LOS velocity, microturbulence). The right panel in Figure \ref{fig:rfs} shows the optical depth ranges where the selected lines in the \irissqp\ are roughly sensitive to changes in the temperature. As mentioned above, these ranges summarize the general behavior of the selected lines for most of the observed solar features in the database. They should be understood as a reference of where the lines are mostly sensitive to changes in the thermodynamics. 

\subsection{Multi-line representative profiles}
Now that we have selected a set of spectral lines that is sensitive to the thermodynamics from the top of the chromosphere to the bottom of the photosphere, the next steps is to prepare them for clustering.
The need for clustering our data is due to the computational resources needed to recover the physical information encoded in the spectral lines, specifically when we are working with the
\cii\ and the \mgii\ lines. As we have already mentioned the inversion of the latter requires consideration of non-LTE and partial frequency redistribution of the scattered photons. Therefore, we follow the same approach introduced by \cite{SainzDalda19}, which was extended by \cite{Woods21} for the multi-line case of \cii\ lines and \mgii\ (including \mguvtt\ lines), and more recently by \cite{SainzDalda22b} for the multi-line case of \cii\ lines, \mgii\ and all the \mguv\ lines.
Now, in addition to the set of chromospheric lines used by the latter authors we include 6 photospheric lines. 
Because we are now simultaneously considering lines sensitive to the thermodynamics at different optical depths we may have RPs which share the same chomospheric physical conditions but different photosperic thermodynamic conditions, or the other way around. 
In other words, identical chromospheric profiles may be associated with photospheric profiles that are quite different, or vice versa. We must consider this degeneracy between the chromospheric set of lines and the photospheric set of lines when clustering them. Thus, compared to inversions that are based on fewer lines, we need to consider a larger number of clusters and RPs to capture properly that degeneracy. We have determined that 320 clusters per data set are able to represent properly most of the data sets considered in our study within the hardware limitations we have.

In this investigation, since we are interested in obtaining thermodynamic models for the ARs we have stratified the data before clustering. 
That means, we have first identified those locations where the profiles belong to the umbra (or pore), the penumbra, the plage,
and the quiet Sun. These four regions are determined by using different intensity thresholds in the reconstructed intensity map
at $2810.58~$\AA of the data set to be clustered. The 3 first panels in the top row of Figure \ref{fig:example_1} show the
intensity spectroheliogram or map for an AR at \mgiik, Ti \small{2} 2785.46 \AA\ and at 2832.04 \AA\ (photospheric continuum)
respectively. The fourth panel shows the areas identified as umbra, penumbra, plage, and quiet sun. Once these areas
are determined, we impose the number of clusters for each region to be clustered. The number of clusters for each region is
based on the average area covered by these regions and the variability of different physical conditions in these areas. For
instance, the quiet Sun occupies more area in most of the data selected compared to the penumbra. On the other hand, the physical conditions are likely more variable in the latter, although that is not necessary true in the chromosphere. We have attempted to find a trade-off between the hardware limitations (to consider at most 320 RPs per data set) and to recover a meaningful representation of the physics of the selected areas in each data set. The number of clusters for the umbra, penumbra, plage and quiet-sun are 30, 50, 80, 160
respectively. If there is no umbra, i.e., no umbra has been detected under the threshold level established, then the features
detected are usually pores, orphan-penumbras, or naked-sunspots. In this case, the number of clusters for the penumbra are
incremented up to 80, and the associated RPs are labeled as 'pore-like'. If there is neither umbra nor penumbra, the number of
clusters for the plage and quiet Sun are 144 and 176 respectively. In total, the \irissqp\ database has $2,280$ RPs associated
with the umbra, $800$ RPs with the pore-like features, $3,800$ RPs with the penumbra, $12,640$ with plage, and $20,800$ RPs with quiet Sun. 
To cluster the original IRIS profiles we created a {\it joint-profile} by cropping the selected lines and concatenating them together. These new profiles are scaled between 0 and 1, then they are clustered with the number of clusters determined by the criteria mentioned above. After the clusters are defined, we calculated the RPs (centroids) as the average of the original profiles, i.e. un-cropped, and associated 
with the elements (labels) of each cluster. At this point, the RPs preserve the original spectral sampling and the original spectral range of the selected lines. They are ready for the inversion. 

\subsection{Inversion of the Multi-Line Representative Profiles}

We have used the multi-atom STockholm inversion Code \citep[STiC][]{delaCruzRodriguez16,delaCruzRodriguez19} to invert the RPs of our selected data. STiC\footnote{STiC is available at \href{https://github.com/jaimedelacruz/stic/}{https://github.com/jaimedelacruz/stic/}.}
is an MPI-parallel non-LTE inversion code that utilises a modified version of RH \citep{Uitenbroek01} to solve the atomic population densities assuming statistical equilibrium and plane-parallel geometry and it allows including partial frequency redistribution effects of scattered photons \citep{Leenaarts12}. The radiative transport equation is solved using cubic Bezier solvers \citep{delaCruzRodriguez13a}. The inversion engine of STiC includes an equation of state extracted from the SME code \citep{Piskunov17}.

An important step in the multi-line inversion is needed to properly handle the strong intensity differences between the various lines. 
This is particularly important for the \cii\ lines since their 
intensity is much  than the \mgii, the \mguv\ lines, and the nearby photospheric lines. Therefore, for the inversion we must scale them with respect to the other lines. We have also considered this between the \mguvtt\ lines with respect to the \mgii\ lines, and the photospheric lines with respect to the \mgii\ lines. The strategy is as follows: we calculate the most frequent value of the intensity in a spectroheliogram map in the \mgii\ $k_3$, and we scale or {\it weight} the \cii\ lines to that value. That means, we try to scale the \cii\ lines to the same relative intensity level as the \mgii\ lines. Note that a somewhat similar strategy was used by \cite{SainzDalda22b}. However, in that case, because the authors were investigating the thermodynamics of the chromosphere during the maximum of a X1.0-class flare,
the ratio between the maximum intensity in the \mgii\ lines and the \cii\ lines varies from $\approx 100$ in
the non-flaring region to just $\approx 10$ in the ribbons, both in the same FoV or dataset. Therefore, the authors decided to stratify the data considering different populations of profiles with different intensity. As a consequence, different weights for the \cii\ lines with respect to the \mgii\ lines were used for different regions in a data set.The need for this scaling (or weighting) is due the different behavior of the lines in different solar features, with the most extreme differences occurring between the non-flaring and flaring regions. The data sets considered in this study do not have such a strong variation of intensity between the \cii\ and \mgii\ lines in different regions within the same field of view. Therefore, only one weight is used for the \cii\ lines for the full data set. For the \mguvo\ line, the \mguvtt\ lines, and the photospheric lines we used a slightly different strategy. For these lines we obtained the weights with respect to the \mgii\ lines in an empirical fashion, after we verified that using the strategy of the most frequent value of the \mgii\ $k_3$ map did not provide a 
significant advantage. In other words, we found that for these lines, on average, the relative intensity with respect to that of the \mgii\ lines for different data sets has a small standard deviation. Thus, for the sake of simplicity, we have used the same combination of weights (with respect to \mgii)
for these lines. For a dataset, STiC only accepts a  wavelength-dependent noise and weight for all the profiles. For simplicity, the noise is the same averaged noise (standard deviation in a spectral range) at all the wavelengths, while the different weights are given for different  spectral positions (usuallly ranges or windows). That means, we use  an unique value of the $w_{i}/\sigma$ per data set in the {\it merit function} used to quantify the quality of the fit between the observed $I(\lambda_i)^{obs}$, and the inverted profile $I(\lambda_i,\mathbf{M})^{syn}$, that is:

\begin{equation}\label{eq:chi2}
\chi^{2} = \frac{1}{\nu}\sum_{i=0}^{q}{(I(\lambda_i)^{obs} - I(\lambda_i, \mathbf{M})^{syn})^{2}\frac{w_{i}^2}{\sigma_{i}^2}} 
\end{equation}

with $i = 0,..., q$ the sampled wavelengths, $w_{i}$ their weights, $\sigma_{i}$ the uncertainties of the observation (e.g. photon noise),
 and $\nu$ the number of spectral samples. Note that we assume, for simplicity, that $\sigma_i$ is the same for all $i$, as mentioned above.
We have used 3 cycles with different number of nodes for the physical variables in the model. These values are detailed in the table \ref{table:inversion}.
We used the FALC quiet-sun model \citep{Fontenla93} as the initial guess model for all the inversions.

\begin{table}[h!]
	\begin{center}
		\begin{tabular}{cccc}
			No. Cycle	&	1 & 2 &  3 \\
			\hline
			$T$  &  4&  7 &  9  \\
			$v_{turb}$	&  2 & 4 & 8 \\
			$v_{los}$	& 2 & 4 & 8 \\
		\end{tabular}
		\caption{Number of nodes in each cycle for the thermodynamics variables considered during the inversions.}\label{table:inversion}
	\end{center}
\end{table}

\section{Results}\label{sec:results}

\begin{figure*}
    \centering
    \includegraphics[width=.95\textwidth]{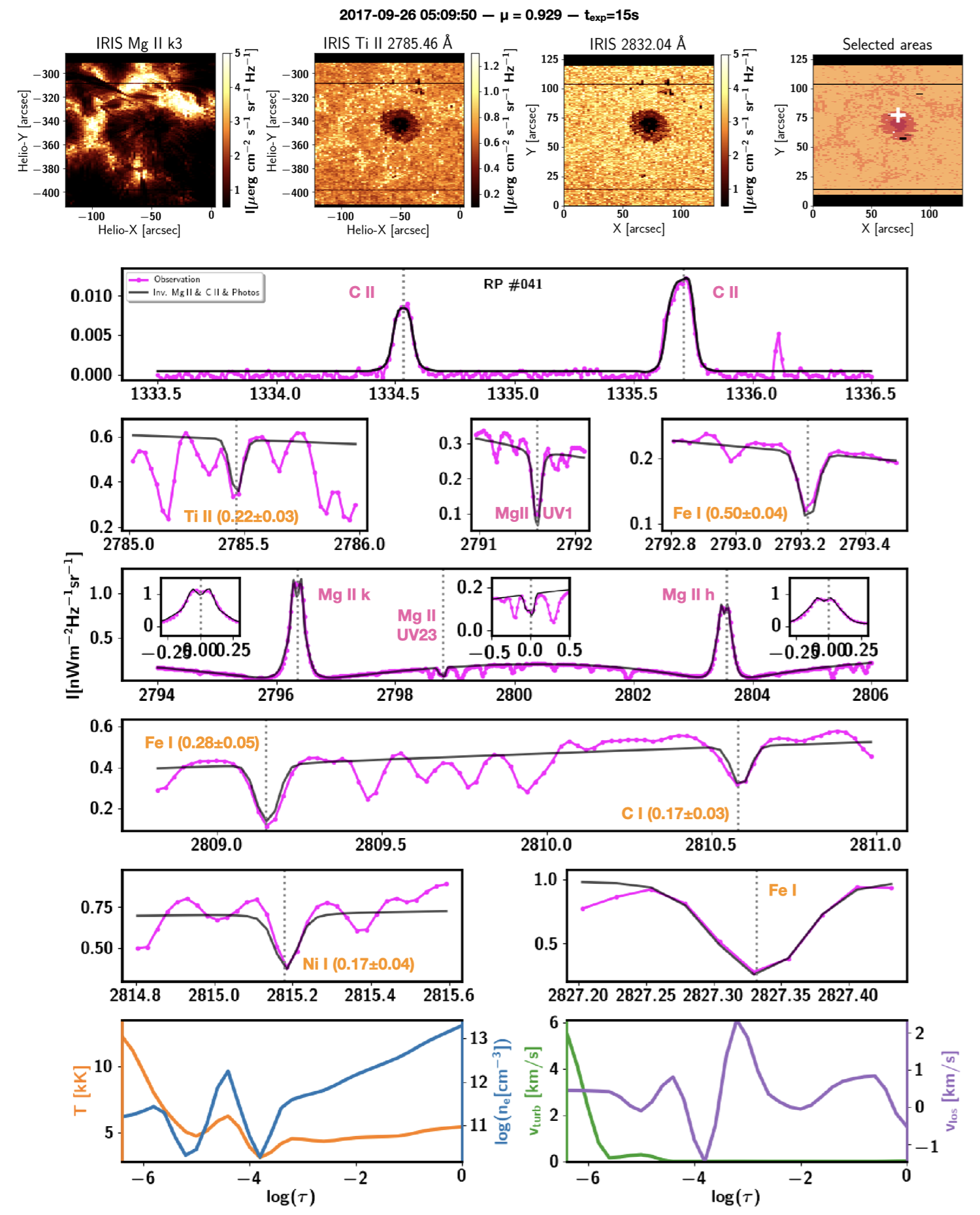}
    \caption{First row: Spectroheliogram map for \mgii\ $k_3$, the \ion{Ti}{1} 2785.46 \AA, the photospheric continuum at 2832.04 \AA, and a map of the various types of regions that were identified (see text). The other rows are results from an inversion for the location marked with a white cross in the ``{\it Selected areas}'' map in the right of the top row. From the second to the sixth row, from left to right, the observed (dotted-solid in fuchsia) and the inverted profile (black) for the \cii\ lines, \ion{Ti}{1} 2785.46 \AA, \mguvo\, \ion{Fe}{1} 2793.22 \AA, \mgii\ lines - including \mguvtt-, \ion{Fe}{1} 2809.15 \AA, \ion{C}{1} 2810.58 \AA, \ion{Ni}{1} 2815.18 \AA, and \ion{Fe}{1} 2827.33 \AA. The bottom row shows the temperature ($T$, orange), the logarithm of the electron density (\nne, violet),  velocity of turbulent motions or micro-turbulent (\vturb, green) and the line-of-sight (\vlos, blue). Both the inverted profile and the representative model atmosphere (RMA) are included in the \irissqp, and they are both labeled as {\tt penumbra} in the database.}
    \label{fig:example_1}
\end{figure*}

\begin{figure*}
    \centering
    \includegraphics[width=.95\textwidth]{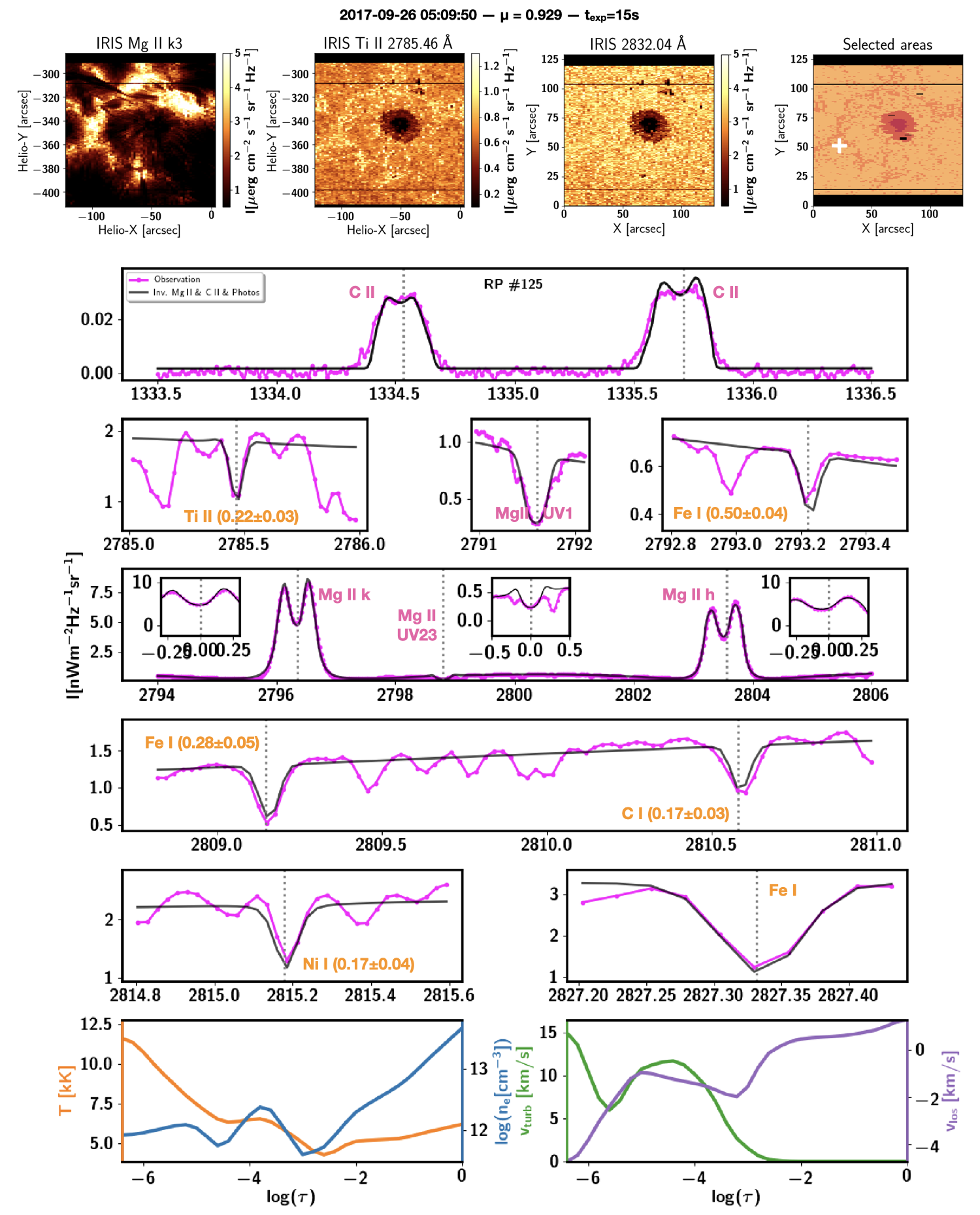}
    \caption{Same as Figure \ref{fig:example_1} for a multi-line RP and RMA in the outer plage (marked with a white cross in the ``Selected areas" panel). Both the inverted profile and eh representative model atmosphere (RMA) are included in the \irissqp, and they are both labeled as '{\tt plage}' in the database.}\label{fig:example_2}
\end{figure*}

Figures \ref{fig:example_1} and \ref{fig:example_2} show  the inversion of a multi-line RP in the penumbra and the outer part of the plage of the NOAA 12681, respectively, both on 2017-09-26 at 05:09:50UT. In addition to the observed RP (in fuchsia) and its corresponding inverted, synthetic RP (best fit obtained by STiC, in black), the temperature ($T$), line-of-sight velocity (\vlos), the velocity of turbulent motions or 
micro-turbulence (\vturb), and the logarithm of the electron density (\nne) are shown in the last row of these figures. The rest position of the lines is indicated with a vertical dashed line.  The inset plots in the fourth row show the core of the \mgii\ and \mguvtt\ lines.

\begin{figure*}
    \centering
    \includegraphics[width=1\textwidth]{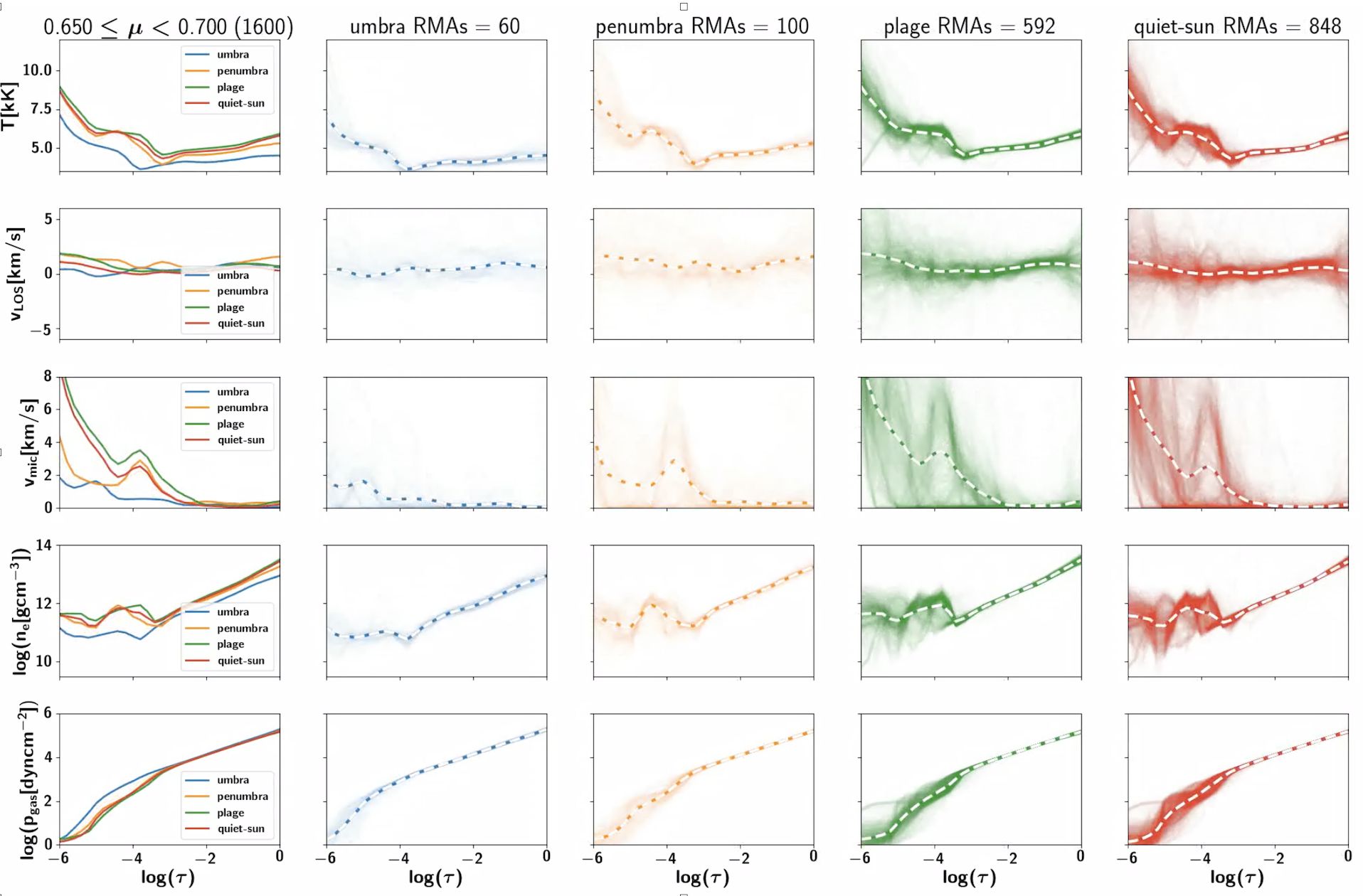}
    \caption{Using \irissqp\ data base to find and visualize the different models of umbra, penumbra, plage and quiet-sun at a given interval of $\mu=cos(\theta)$, with $\theta$ the heliocentric angle.}\label{fig:example_db}
\end{figure*}

As a result of the strategy presented in this article, we
have an indexed, labeled, relational data base. Thus, each
inverted synthetic RP has an associated
representative model atmosphere (RMA), and the following
meta-data: the IRIS filename and the solar feature (``{\tt umbra}'',
``{\tt penumbra}'', ``{\tt pore-like}'', ``{\tt plage}'', and ``{\tt quiet-sun}'') where the
RP was clustered, the position on the Sun of the original observed profiles associated with the RP's cluster, the
exposure time, and the observation time. Because
information is attached to the RP-RMA pair in \irissqp\, one can 
search for RMAs for a given solar feature (e.g. '{\tt plage}'), at a given time in the solar cycle (e.g. '{\tt 2014-01-01}' $<$ {\it date\_obs} $<$ '{\tt 2014-05-31}'), and at 
given position on the Sun. Figure \ref{fig:example_db} shows
the case of a search of the RMAs for all the solar features
located at $0.65 < \mu  < 0.70$. As one would expect,
the RMAs show, for a given physical parameter, some variation both within the same solar feature  (e.g., the \vturb\ for
the RMAs in the plage), and between solar features (e.g umbra
versus penumbra). In this example search, only 2 umbras (30
RMA/umbra) and 2 penumbras (50 RMA/penumbra) were observed
and included in the data base. In fact, these 2 umbras and 2
penumbras observations belong to the same AR, and for that
reason the dispersion between the RMAs for the different
physical variables is rather small in these solar features.
Several plages and "quiet-sun" belonging to different ARs are
however found in this search, and for that reason their
associated RMAs show a larger variability. We remind the reader that the term "quiet-sun" here refers to regions within an active region that are not plage, umbra, pores, or penumbra. It is not evident that such regions are, in fact, identical to quiet Sun regions that are not part of an active region. This is because the overlying large-scale magnetic field topology and the associated canopy can impact the chromosphere of these regions, which are at the photospheric level perhaps more quiet-Sun like.

The quality of the inversion is not always easy to evaluate from the $\chi^2$ value in the profiles of \irissqp. This is because of the behavior of the Euclidean distance when applied to problems with high dimensions, which is the case of our profiles. In many cases, a bad fit in the core of the \mgii\ lines penalizes too much the value of $\chi^2$ (see section 3 of \citealt{SainzDalda22a} for a more detailed discussion). In addition, we are now using multiple lines. That increases the possibility of having a bad $\chi^2$ value when one or a few lines are badly fitted, but the rest are well fitted. Of course, when all the lines are well fitted, the $\chi^2$ is good (i.e., low). The question is what to do when several lines are well fitted and others are not (i.e., with a resulting high value for $\chi^2$). As we will discuss later, that is not a big problem, since we have verified that in a large majority the profiles are well fitted. Even for the minority of RPs where some lines are badly fitted, there are usually several other lines that are well fitted.

\section{Discussion}\label{sec:discussion}
In the example of the \irissqp\ data base shown in Figure \ref{fig:example_1}, the inverted profiles fit the observed
profiles very well. Therefore, we can be confident with the representative model
atmosphere (RMA) obtained, and because the fit is very good for all the
lines, the uncertainty of the physical values should be rather small at all the optical depth ranges
shown in the right panel of Figure \ref{fig:rfs}, i.e., where the RFs are large.
What if one or more lines are not fitting well the observed profile? A
very picky reader might consider that is the case, for instance, for the
lines \ion{Fe}{1} 2793.22 \AA, \ion{Fe}{1} 2809.15 \AA, \ion{C}{1}
2810.58 \AA, and \ion{Ni}{1} 2815.18 \AA. In this case, the reader
should consider the values at the optical depths $-3<$ \ltau $<-1$ with some
caution, since only \ion{Fe}{1} 2827.33 \AA is fitting the observed
profile very well. Which impact do the lines that do not fit the observations well have in the \irissqp?
Strictly speaking, none. The \irissqp\ database is a relational data
base between model atmospheres and their synthetic profiles. The physics
encoded in the latter are strictly due to the values in the model
atmosphere and the considerations made to solve the radiative transfer
equation. Practically speaking, it has an impact if the fits were
consistently bad - which is not the case. 

The \irissq\ can be used in
different ways: as a set of reference model atmospheres and profiles for the various types of features within ARs; as constraints for numerical models; as a look-up
table for speeding inversions problems; or as a source atmosphere to 
synthesize spectral lines both in the chromosphere and the photosphere, including full-Stokes polarimetric profiles by considering magnetic field information. 
In all these cases, we want to have a better representation of the
thermodynamics in the low solar atmosphere, and in that sense the better
the fits are, the better that representation is. However, those
synthetics profiles that show some poorer fits with respect to the observed RP very likely still {\it occur} in the solar atmosphere. Therefore, since a very large number of the inversions calculated to build the \irissqp\ data base are showing very good fits, we consider this database likely the most accurate comprehensive set of
stratified-depth models and profiles in the chromosphere and the
photosphere.

\section{Conclusions}

Thanks to the extraordinary multi-line capabilities of IRIS, we have created a database of 40,320 synthetic representative profiles and their corresponding representative model atmospheres that convey the essential thermodynamic information of active regions from the bottom of the photosphere to the top of the chromosphere. 

Thanks to the state-of-the-art multi-line multi-atom STiC inversion code, we have been able to recover the physical information encoded in 6 chromospheric lines and 6 photospheric lines. Using the well-known k-means technique we have clustered multi-line IRIS spectral data. This helps us to overcome the task of inverting all the profiles considered in this work, which would require very significant computational resources ($\approx 242,000~CPU-hours$).

\irissqp\ is an unique data base, obtained
systematically from an unprecedented combination of a
large number spectral lines sensitive to the
thermodynamics in the low solar atmosphere. We will
continue maximizing the observational capabilities of
IRIS, considering more lines and applying improved
inversion methods and advance machine learning
techniques. \irissqp\ has a high potential to be used
in different ways to gain knowledge in the low solar
atmosphere. \irissqp, as well as \irissq\ already is,
will be available to the public in different formats,
both for IDL and Python. We encourage the community to
exploit \irissqp\ to solve the physics of the ARs.

\acknowledgments
IRIS is a NASA small explorer mission developed and operated by LMSAL with mission operations executed at NASA Ames Research center and major contributions to downlink communications funded by ESA and the Norwegian Space Agency. This work was supported by NASA contract NNG09FA40C (IRIS).
A. Agrawal was supported by the Lockheed-Martin-Palo Alto Unified School District internship program. The authors are grateful to Mats Carlsson and Tiago Pereira for providing the C atom model, and to Jaime de la Cruz Rodríguez for his helpful comments on the inversion of photospheric lines.
\software{\href{https://iris.lmsal.com/iris2/}{\irissq}}.

\bibliographystyle{yahapj}
\bibliography{allbib,others}

\end{document}